\documentclass[sigconf]{acmart}

\usepackage{subfigure}
\usepackage{makecell}
\usepackage{url}
\usepackage{xcolor}
\usepackage{xspace}
\usepackage{hyperref}
\usepackage{enumitem}
\usepackage{cleveref}
\usepackage{amsfonts}
\usepackage{multirow}
\usepackage{adjustbox}
\usepackage{colortbl}
\usepackage[ruled,vlined]{algorithm2e}
\usepackage{balance}

\usepackage{acronym}

\acrodef{ZKP}[ZKP]{Zero-knowledge proof}

\acrodef{FL}[FL]{Federated Learning}

\newcommand{\AIArena}{\textsc{AIArena}\xspace}

\copyrightyear{2025}
\acmYear{2025}
\setcopyright{cc}
\setcctype{by}
\acmConference[WWW Companion '25]{Companion Proceedings of the ACM Web Conference 2025}{April 28-May 2, 2025}{Sydney, NSW, Australia}
\acmBooktitle{Companion Proceedings of the ACM Web Conference 2025 (WWW Companion '25), April 28-May 2, 2025, Sydney, NSW, Australia}
\acmDOI{10.1145/3701716.3715484}
\acmISBN{979-8-4007-1331-6/2025/04}

\begin{document}
\begin{abstract}
The rapid advancement of AI has underscored critical challenges in its development and implementation, largely due to centralized control by a few major corporations. This concentration of power intensifies biases within AI models, resulting from inadequate governance and oversight mechanisms. Additionally, it limits public involvement and heightens concerns about the integrity of model generation. Such monopolistic control over data and AI outputs threatens both innovation and fair data usage, as users inadvertently contribute data that primarily benefits these corporations.

In this work, we propose \AIArena, a blockchain-based decentralized AI training platform designed to democratize AI development and alignment through on-chain incentive mechanisms. \AIArena fosters an open and collaborative environment where participants can contribute models and computing resources. Its on-chain consensus mechanism ensures fair rewards for participants based on their contributions. We instantiate and implement \AIArena on the public Base blockchain Sepolia testnet, and the evaluation results demonstrate the feasibility of \AIArena in real-world applications. Our project page is available at \url{https://train.flock.io/explore}.
% \href{https://train.flock.io/explore}{here}.

\end{abstract}

\title{\textsc{AIArena}: A Blockchain-Based Decentralized AI Training Platform}

% \author{Zhipeng Wang${^\dagger}$, Rui Sun${^\ddagger}$, Elizabeth Lui${^\mathsection}$, Tuo Zhou${^\mathparagraph}$, Yizhe Wen${^\mathsection}$, 
%  and  Jiahao Sun${^{\mathsection} }$}
% \affiliation{%
%   \institution{${^\dagger}$Imperial College London, ${^\ddagger}$Newcastle University, ${^ \mathsection}$FLock.io, ${^\mathparagraph}$The University of Hong Kong}
%   \country{}
% }

\author{Zhipeng Wang}
\affiliation{
  \institution{Imperial College London}
  \city{London}
  \country{United Kingdom}
}
\email{zhipeng.wang20@imperial.ac.uk}

\author{Rui Sun}
\affiliation{
  \institution{Newcastle University}
  \city{Newcastle upon Tyne}
  \country{United Kingdom}
}
\email{rui.sun@newcastle.ac.uk}

\author{Elizabeth Lui}
\affiliation{
  \institution{FLock.io}
  \city{London}
  \country{United Kingdom}
}
\email{elizabeth@flock.io}

\author{Tuo Zhou}
\affiliation{
  \institution{The University of Hong Kong}
  \city{Hong Kong}
  \country{China}
}
\email{zhoutuo@connect.hku.hk}

\author{Yizhe Wen}
\affiliation{
  \institution{FLock.io}
  \city{London}
  \country{United Kingdom}
}
\email{yizhe@flock.io}

\author{Jiahao Sun}
\affiliation{
  \institution{FLock.io}
  \city{London}
  \country{United Kingdom}
}
\email{sun@flock.io}

% https://conferences.sigcomm.org/imc/2024/#:~:text=The%202024%20Internet%20Measurement%20Conference,Internet%20Measurement%20Workshops%20and%20Conferences.

% \renewcommand{\shortauthors}{Wang et al.}

\renewcommand{\shortauthors}{Zhipeng Wang et al.}




\maketitle

%------------------------------------------------------%
%--------------------- main ---------------------------%

\section{Introduction}

The growth of artificial intelligence (AI) has led to groundbreaking applications in many fields. However, it also brings challenges, especially because a few large companies control most of its development and use. This concentration of power increases bias in AI systems~\cite{challen2019artificial} and reduces public involvement in important decisions. This lack of transparency can exacerbate issues such as misuse of AI technologies or unethical practices, as there is little external oversight in the AI centralized systems. Such centralized control also slows down innovation in AI~\cite{Etzioni2016Democratizing, montes2019distributed} and leads to the unfair use of user data, which mainly benefits these companies. These compounded challenges underline the urgent need for a shift to decentralized AI (DeAI)~\cite{wang2024sok}, where control and access are distributed across a broader and more inclusive range of participants.

 Blockchain's decentralized nature can help address the above challenges by allowing multiple participants to collaborate on AI development without relying on a central authority~\cite{chen2018machine, dong2024defending}. Smart contracts~\cite{zou2019smart} can automate processes such as distributing rewards or verifying contributions, fostering fairness and efficiency. Blockchain also enhances data integrity, as participants can verify the source and ownership of data, preventing unauthorized use and ensuring compliance with ethical standards~\cite{salah2019blockchain}.

\begin{figure}[t]
\centering
\includegraphics[width=\columnwidth]{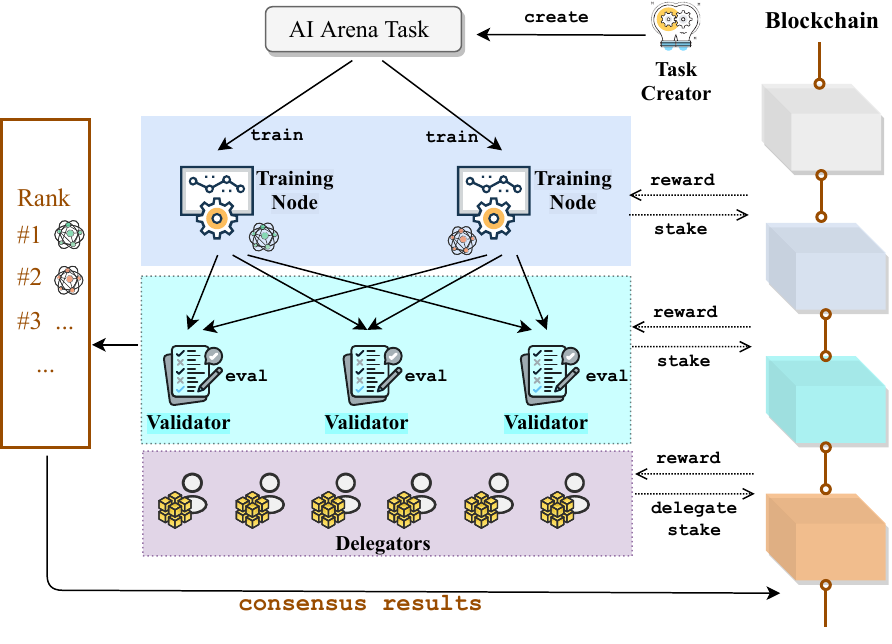}
% \vspace{-8mm}

\caption{Overview of \AIArena, a blockchain-based decentralized AI training system.}
\label{fig:AIArena-system}
\end{figure}

In this work, we propose \AIArena, a blockchain-based platform designed to decentralize AI training in an open and collaborative environment, where participants can contribute models and computing power. Specifically, \AIArena leverages on-chain consensus mechanisms to ensure that only valid contributions are rewarded, while incentivization models encourage active and meaningful participation to prevent issues such as free-riding or fake identities.

\noindent \textbf{Contributions.} Our contributions are summarized as follows:

% \vspace{-5mm}

\begin{itemize}[leftmargin = *]
    \item We propose \AIArena, a blockchain-based decentralized platform for AI training. \AIArena enables the AI model to be optimized directly on users’ devices using their own data or publicly available datasets, and finally derives the optimal model based on blockchain consensus. By leveraging blockchain, \AIArena ensures sustained machine learning (ML) contributor (i.e., training node and validator) engagement and provides fair rewards based on their contributions in enhancing models.

    \item We introduce a delegation stake design for \AIArena, enabling users with limited computational resources to participate in model generation by delegating their stake to preferred ML participants. This design encourages broader involvement in the ML model generation process, further promoting decentralization. 
    
    \item We instantiate and implement \AIArena on the public Base blockchain Sepolia testnet\footnote{ \href{https://sepolia.basescan.org/address/0x7b6bde1d173eb288f390ff36e21801f42c4d8d91}{https://sepolia.basescan.org/address/0x7b6bde1d173eb288f390ff36e21801f42c4d8d91}}. Over an approximate $7$-month period, \AIArena has engaged $603$ training nodes, $1{,}051$ validators, and $63{,}265$ delegators, collaboratively generating $18{,}656$ models across $16$ training tasks. These results highlight the practical feasibility of \AIArena in real-world applications.

    \item We evaluate three popular tasks performed by  \AIArena contributors. Their models not only outperform baseline models but also exceed the performance of larger state-of-the-art (SOTA) models. Notably, in the code co-creation task, our contributors collaboratively curated the largest existing Move code dataset, along with their corresponding code instructions and comments. 
    
\end{itemize}

\section{\AIArena System Design}

\subsection{Participants}

Fig.~\ref{fig:AIArena-system} shows the overview of \AIArena system. There are various categories of participants in the \AIArena system:

\begin{itemize}
    [leftmargin = *]
    \item \textbf{Task Creators:} Task creators are responsible for defining training tasks and their specific requirements, including selecting and designing appropriate learning algorithms for model training and validation. To promote decentralization, the created tasks can be reviewed and verified by the stakeholders within \AIArena.

    % Initially, only FLock's core team could create tasks, but this will expand to the broader ecosystem. Tasks verified by the community-led DAO are eligible for FML daily emission as rewards, while tasks created permission-lessly must be self-funded by the creator, balancing innovation with quality control.

    \item \textbf{Training Nodes:} Given a created task, training nodes are 
    responsible for training the model on their service using public data. To participate, training nodes must stake blockchain assets or tokens to establish their eligibility. Training nodes will receive rewards based on their stake amount and performance.

    \item \textbf{Validators:} Validators are responsible for assessing the work completed by training nodes and submitting validation scores that influence the distribution of rewards. To participate, validators must stake tokens, which not only allows them to validate assigned tasks but also ensures fair task allocation.

    \item \textbf{Delegators:} Delegators contribute to the \AIArena system by supporting the staking process of other participants without directly engaging in task training or validation. They stake tokens on behalf of other participants, boosting the delegatees’ capacity to earn greater rewards. In return, delegators share in the rewards earned by their delegatees, according to predefined algorithms that reflect their staked contributions.

\end{itemize}

\subsection{Training and Validation}

% \subsubsection{Training Node and Validator Selection}
% All training nodes and validators must first stake tokens in the system to qualify for performing task training or validation. To ensure fairness and efficiency, a rate-limiting mechanism is implemented to control how often participants can access validation opportunities for a given task. T

% he probability of being selected to validate a task submission increases with the participant's stake. However, the increase in validation frequency diminishes progressively as the staking amount grows, ensuring a balanced and equitable system.

\subsubsection{Training}

We consider the dataset held by the training node, \(\mathcal{D}_{\text{local}}\), which contains locally sourced data samples, comprising feature set \(X\) and label set \(Y\), with each sample \(x_i \in X\) corresponding to a label \(y_i \in Y\). We define a predictive model \(f\), aiming to learn patterns within \(\mathcal{D}\) such that \(f(x_i) \approx y_i\).

To quantify the prediction metric, e.g., accuracy, the task trainer will introduce a loss function \(L(f(x_i), y_i)\), assessing the discrepancy between predictions \(f(x_i)\) and actual labels \(y_i\). A generic expression for this function is: $L = \frac{1}{N} \sum_{i=1}^{N} l(f(x_i), y_i)$, where \(N\) denotes the total sample count, and \(l\) signifies a problem-specific loss function, e.g., mean squared error or cross-entropy loss.

The optimization goal is to adjust the model parameters \(\theta\) to minimize \(L\), typically through algorithms such as  gradient descent: $\theta_{\text{new}} = \theta_{\text{old}} - \eta \nabla_\theta L$, where \(\eta\) represents the learning rate, and \(\nabla_\theta L\) the gradient of \(L\) with respect to \(\theta\). Utilizing the aggregated dataset \(\mathcal{D}\), parameter \(\theta\) is iteratively updated to reduce \(L\), consequently improving the model's predictive accuracy. This optimization process is conducted over a predefined number of epochs \(E\), each epoch consisting of a complete pass through the entire dataset \(\mathcal{D}\).

\subsubsection{Validation}
Consider a selected group of validators, denoted as $V_j \in V$, each equipped with the evaluation dataset $\mathcal{D}_{\text{eval}}$ from the task creator. This dataset consists of pairs $(x_i, y_i)$, where $x_i$ represents the features of the $i$-th sample, and $y_i$ is the corresponding true label.  The model, trained by designated training nodes, is denoted as $\theta^{task}_p$. The primary objective of $\theta^{task}_p$ is to predict the label $\hat{y}_i$ for each feature vector $x_i$ contained within $\mathcal{D}_{\text{eval}}$. To assess the performance of $\theta^{task}_p$ on $\mathcal{D}_{\text{eval}}$, we use a general evaluation metric denoted by $eval$. We exemplify with accuracy as follows:

\begin{equation}\label{eq:eval_eq}
eval(\theta^{task}_p, \mathcal{D}_{\text{eval}}) = \frac{1}{|\mathcal{D}_{\text{eval}}|} \sum_{(x_i, y_i) \in \mathcal{D}_{\text{eval}}} \mathbf{1}(\hat{y}_i = y_i)
\end{equation}

Here, $\mathbf{1}$ represents the indicator function that returns 1 if the predicted label $\hat{y}_i$ matches the true label $y_i$, and 0 otherwise. The function $|\mathcal{D}_{\text{eval}}|$ denotes the total number of samples within the evaluation dataset.
Each predicted label $\hat{y}_i$ from the model $\theta^{task}_p$ is compared against its corresponding true label $y_i$ within $\mathcal{D}_{\text{eval}}$. The calculated metric result (e.g., accuracy) serves as a quantifiable measure of $\theta^{task}_p$'s effectiveness across the evaluation dataset.

\subsection{Consensus and Reward Distribution}

We assume there are $n$ submissions $(O_1, \ldots, O_n)$ from $n$ training nodes with stakes $(t_1, \ldots, t_n)$, and $m$ validators $(V_1, \ldots, V_m)$ with stakes $(s_1, \ldots, s_m)$. Each validator $V_j (1\leq j\leq m)$ evaluates the $n$ models submitted by the training nodes, producing a score vector $\vec{r}_j = (r_{j1}, \ldots, r_{jn})$. These scores reflect the performance of each model according to predefined criteria as shown in Eq.~\ref{eq:eval_eq}. 

\subsubsection{Reward Distribution within One Task}
Within a single task, the reward distribution between training nodes and validators is determined based on their relative stake amounts. Let the total daily reward allocated to a task be denoted as \( R_0 \). The total rewards for training nodes are: $R_0 \cdot \left( \gamma + (1 - 2\gamma) \cdot \frac{\sum_{i=1}^{n} t_i}{\sum_{i=1}^{n} t_i + \sum_{j=1}^{m} s_j} \right)$. Similarly, the total rewards for validators are: $
R_0 \cdot \left( \gamma + (1 - 2\gamma) \cdot \frac{\sum_{j=1}^{m} s_j}{\sum_{i=1}^{n} t_i + \sum_{j=1}^{m} s_j} \right)
$. The parameter \( \gamma \) controls the split rewards, defining the balance between fixed and stake-dependent reward components.

\subsubsection{Reward for Training Nodes}
The final scores of the submitted model are determined through a weighted aggregation: $
    \vec{r} = \left(\frac{\sum_{j}{r_{j1}\cdot s_j}}{\sum_{j}{s_j}}, \ldots, \frac{\sum_{j}{r_{jn}\cdot s_j}}{\sum_{j}{s_j}}\right)
    $. This means that the evaluations of validators with higher stakes have a larger impact on the final outcome. We then compute the following geometry series:
    $g_k =  \frac{1 - q}{1 - q^m} \cdot q^{k-1}$, 
    in which $k$ denotes a given training node's rank amongst its peers in the same task, whereas $q$ is the common ratio of the geometric series and $m$ is the number of training nodes in a given task.

% Participants in the FLock system, such as training nodes and validators, are required to contribute computing or storage resources to complete model training and validation in order to receive rewards. This means that the value of the FML token will, at a minimum, correspond to the value of the resources consumed during these processes.
 We finally compute the total rewards allocated for the training nodes as well as their delegators, which is based on the quality of their submission and their total amount of stake: $f_i(g_i, t_i) = \frac{g_{i} \cdot t_i^{\alpha_t}}{\sum_{k=1}^{n} g_{k} \cdot t_k^{\alpha_t}}$, where $t_i$ the total stake amount from the training node $i$ as well as its respective delegators. $\alpha_t$ is a system parameter that determines the influence of the stake on the reward distribution.

% \begin{equation} \label{eq: training_reward}
%     f_i(g_i, t_i) = \frac{g_{i} \cdot t_i^{\alpha_t}}{\sum_{k=1}^{n} g_{k} \cdot t_k^{\alpha_t}}
% \end{equation}

If a training node $i$'s stake in the task is $t_n$ and stakes delegated to training node $i$ is $t_d$, i.e.,  $t_i$ = $t_n$ + $t_d$, then the actual reward for training node $i$ is $f_i \cdot \left(\sigma + (1-\sigma) \cdot \frac{t_n}{t_n+t_d}\right)$. $\sigma$ is the reward ratio set by the training node, which determines the ratio of rewards shared between the training node $i$  and its respective delegators.

    % \item Security and integrity of the evaluation process are paramount. Validators are incentivised to participate honestly through mechanisms such as slashing, where those who behave maliciously or negligently have their stakes reduced.
    % \item The system should also be safeguarded against collusion among validators or attempts to manipulate the evaluation process. Techniques such as advanced cryptography, reputation systems, and random validator selection can enhance trust and decentralisation in the evaluation framework.

\subsubsection{Reward for Validators}

For each validator $V_j$, we compute the distances between their score and the final aggregated score: $ \vec{\Delta_j} = \left(\Delta_{j1}, ..., \Delta_{jn}\right)
        = \left(\left|{\frac{\sum_j{r_{j1}\cdot s_j}}{\sum_j{s_j}}}-r_{j1}\right|, ..., \left|\frac{\sum_j{r_{jn}\cdot s_j}}{\sum_j{s_j}} - r_{jn}\right|\right) $.

% \begin{equation*}
%     \begin{aligned}
%          \vec{\Delta_j} &= \left(\Delta_{j1}, ..., \Delta_{jn}\right)\\
%          &= \left(\left|{\frac{\sum_j{r_{j1}\cdot s_j}}{\sum_j{s_j}}}-r_{j1}\right|, ..., \left|\frac{\sum_j{r_{jn}\cdot s_j}}{\sum_j{s_j}} - r_{jn}\right|\right) 
%     \end{aligned}
% \end{equation*}
We define a distribution function $f_i$, which satisfies: \emph{(1)}  $f_i(\Delta_{1i}, s_1) + \ldots + f_i(\Delta_{mi}, s_m) = 1$; \emph{(2)} $f_i$  decreases over the distance $\Delta_{ji}$, and \emph{(3)} $f_i$ increases over the stake amount $s_j$.
  To fulfill the three criteria, we can employ a modified version of the Softmax Function: $f_i(\Delta_{ji}, s_j) = \frac{e^{-\lambda_v \Delta_{ji}} \cdot s_j^{\alpha_v}}{\sum_{k=1}^{m} e^{-\lambda_v \Delta_{ki}} \cdot s_k^{\alpha_v}}$. The parameters $\lambda$ and $\alpha$ play  crucial roles. $\lambda_v$ controls the sensitivity of the function to the distance $\Delta_{ji}$. A higher $\lambda_v$ increases the function's sensitivity to score accuracy, emphasizing the importance of precise evaluations. $\alpha_v$ determines the influence of the stake amount $s_j$ on the reward distribution.
  
  % thereby adjusting the weight given to validators' financial contributions.

% \begin{equation}\label{eq: validator-reward}
%     f_i(\Delta_{ji}, s_j) = \frac{e^{-\lambda_v \Delta_{ji}} \cdot s_j^{\alpha_v}}{\sum_{k=1}^{m} e^{-\lambda_v \Delta_{ki}} \cdot s_k^{\alpha_v}}
% \end{equation}

% \begin{itemize}
%     \item {$\lambda_v$}

% \begin{itemize}
%     \item Purpose: Controls the sensitivity of the function to the distance $\Delta_{ji}$. This distance measures the discrepancy between a validator's score and the aggregated score.
%     \item Effect: A higher $\lambda_v$ increases the function's sensitivity to score accuracy, emphasising the importance of precise evaluations.
    
%     \item Selection Criteria: The choice of $\lambda_v$ balances the need to penalise inaccuracies against the goal of rewarding nearly accurate evaluations.
% \end{itemize}

% \item {$\alpha_v$}
% \begin{itemize}
%     \item Purpose: Determines the influence of the stake amount $s_j$ on the reward distribution, thereby adjusting the weight given to validators' financial contributions.
    
%     \item Effect:  Allows for balancing between the importance of validators' financial commitment and their performance accuracy. A higher $\alpha_v$ gives more weight to the stake amount in the reward calculation.
    
%     \item Selection Criteria: Reflects the system's philosophy regarding the stake's importance relative to score accuracy. An $\alpha_v$ of 0 means stake amounts are ignored, while a higher value increases their impact.
% \end{itemize}
% \end{itemize}

If the validator finishes multiple (i.e., $N$) validation tasks, then its reward ratio is $\sum_{i = 1}^{n}{f_i(\Delta_{ji}, s_j)}$. If a validator’s stake in the task is $s_v$, and $s_j$ is its accumulative stake by considering the total delegation amount $s_d$ on this validator, i.e., $s_j = s_v + s_d$, the actual reward ratio for this validator is $\left(\sum_i{f_i(\Delta_{ji}, s_j)}\right) \cdot \left(\sigma + (1-\sigma) \cdot \frac{s_v}{s_v+s_d}\right)$, where $\sigma$ is a parameter set by the delegatee.

\subsection{Delegate Staking}
Delegators may entrust their tokens to participants of their choosing to receive a passive investment income stream. The receivers can thus amplify their stake and rewards. These rewards are shared with the delegators, to attract users who have tokens but lack the technical expertise to perform AI model training or validation.

Specifically, the delegator rewards depend on: \emph{(1)} The quality of the training nodes or validators selected for delegation, and \emph{(2)} The amount of stake delegator has delegated.

{Formally, the the reward for a delegator who delegates to a training node can be calculated as: $f_i \cdot (1-\sigma) \cdot \frac{t_d}{t_n+t_d}$,} whereas $f_i$ refers to the total reward distributed to the training node $i$ and delegator based on the quality of the training node's submission, $t_d$ is the stake amount from this given delegator,  $t_n$ is the stake amount from the training node $i$. Similarly, the reward for a delegator who delegates to a validator can be calculated as: 
{$\left(\sum_i{f_i(\Delta_{ji}, s_j)}\right) \cdot (1-\sigma) \cdot \frac{s_d}{s_v+s_d}$}in which $s_d$ refers to the stake amount from a given delegator and  $s_v$ is the stake amount of the validator the delegator delegated to. 

% $\sigma$ is the reward share ratio pre-determined by the delegate.

% \subsection{Watermarking-based Proof of Training}
% To mitigate the model stealing in the decentralized system, we propose a  proof of training scheme based on watermarking. 
% \begin{itemize} [leftmargin = *]
%     \item When a training node performs the training, they insert the watermarking based on their public blockchain address into the model and submit the model to validators.

% \end{itemize}

\subsection{Various Validation Phases} To address potential attacks, such as lookup attacks and model-stealing attacks by malicious training nodes, \AIArena validators employ diverse validation datasets across different phases. For a task spanning \( x \) days, the process is divided into three phases:

% \vspace{-1mm}
\begin{enumerate}[leftmargin=*]
    \item \textbf{Submission Phase}: This initial phase lasts \( x_0 \) days and provides daily rewards to training nodes based on the validator consensus results for each day. The continuous reward mechanism encourages sustained participation by offering consistent feedback and reinforcing contributors' efforts.

    \item \textbf{Final Validation Phase}: Lasting \( x_1 \) days, this phase utilizes a validation dataset distinct from the one used in the submission phase. This diversification increases the difficulty for malicious actors to exploit predictable validation scenarios, enhancing the system’s robustness.

    \item \textbf{Challenging Phase}: Spanning \( x - x_0 - x_1 \) days, this phase allows any validator to issue a challenge if they suspect a model has been stolen. Training nodes must provide proof of legitimate training, such as proof of learning~\cite{jia2021proof, lan2021proof} or on-chain address-based watermarking~\cite{zhang2020model}. If they fail to do so, the successful challenger receives the rewards allocated to the malicious training node, incentivizing honest participation.
\end{enumerate}
This phased approach ensures a dynamic and secure validation process, mitigating risks associated with malicious training activities.

% the validation dataset used during the initial $x-1$ days differs from that of the final day. These distinct validation datasets are associated with two types of rewards: \textbf{Reward A} for the initial period and \textbf{Reward B} for the final day. This strategic approach enhances security by varying the data against which training nodes are validated, thereby complicating any potential malicious attempts to exploit predictable validation scenarios.

%     \begin{itemize} [leftmargin = *]
%         \item \textbf{Reward A} provides a daily contingent reinforcement schedule, incentivizing participant engagement with tasks through immediate gratification. This continuous reward mechanism fosters sustained participation by providing consistent feedback and reinforcing contributions. Assume the total reward for the training nodes is $R^{AI}_i$, then Reward A of the task is $R^{AI, A}_i = \delta \cdot R^{AI}_i$, where $\delta$ is a configurable system parameter.
        
%         \item \textbf{Reward B}, implements vesting, releasing tokens upon successful task completion within a predetermined lifespan. This incentivizes training nodes to both engage in tasks and ensure their timely and efficient completion. Vesting also functions as a quality control mechanism, promoting focused contributions aligned with task deadlines. Reward B of the task is:
%         $R^{AI, B}_i = (1-\delta) \cdot R^{AI}_i$
%     \end{itemize}

% \vspace{-10mm}

\begin{table*}[t]
    \centering
    \renewcommand\arraystretch{1}
    \resizebox{0.95\linewidth}{!}{
    \begin{tabular}{c|c|c|cccc|cccc|c}
    \toprule
    \multirow{3}{*}{Task ID} & \multirow{3}{*}{Task Name} & \multirow{3}{*}{\makecell{Duration\\ (days)}}  & \multicolumn{4}{c|}{Training Nodes} & \multicolumn{4}{c|}{Validators} &\multirow{3}{*}{Delegators}\\
    \cline{4-11}
    
     &  &    & \multirow{2}{*}{Number} & \multicolumn{3}{c|}{Rewards} & \multirow{2}{*}{Number} & \multicolumn{3}{c|}{Rewards} & \\
    \cline{5-7} \cline{9-11}
    
     &  &  &  \textbf{}  & Mean$\pm$Std & Min & Max &  & Mean$\pm$Std & Min & Max &\\
    \hline
    
     1  & D\&D MBTI Chatbot & 10 &   \cellcolor{gray!30}5 & \cellcolor{blue!10}1879.50$\pm$1614.41 & 11.00 & 3895.53 & \cellcolor{gray!30}6 & \cellcolor{blue!10}708.84$\pm$63.79 & 666.05 & 821.97 & 5 \\
     
     2  & CharacterX - Yae Miko & 14   & 11 & 1033.72$\pm$402.23 & 537.17 & 1745.63 & 9 & 483.30$\pm$306.93 & 33.45 & 930.03 & 3 \\
     3  & AI Friend - Loki & 21 &  39 & 635.06$\pm$621.60 & 59.99 & 2177.59 & 49 & 188.55$\pm$504.51 & 0.00 & 2850.25 & 21 \\
     4  & AI Friend - Hutao & 21   & 38 & 733.90$\pm$588.06 & 124.82 & 1948.08 & 46 & 220.17$\pm$322.86 & 0.00 & 1533.65 & 11 \\
     5  & Farcaster GPT  & 18 &  35 & 334.59$\pm$361.29 & 2.30 & 1240.02 & 34 & 334.13$\pm$326.71 & 0.04 & 877.44 & 46 \\
     
     6  & Multilingual Query Analyzer & 15  & 27 & 423.95$\pm$352.11 & 9.10 & 1263.14 & 48 & 204.57$\pm$311.66 & 0.00 & 1501.77 & 37 \\
     
     7  & Professor Grump & 6 &   13 & 268.50$\pm$151.78 & 91.08 & 543.35 & 38 & 117.47$\pm$133.31 & 1.27 & 510.54 & 17 \\
     8  & FLock GPT & 21 &   31 & 443.31$\pm$674.22 & 8.90 & 2505.02 & 102 & 146.40$\pm$207.77 & 0.02 & 1089.89 & 36 \\
     9  & Text2SQL Agent Model & 28   & 153 & 113.82$\pm$296.88 & 0.00 & 1600.63 & 172 & 95.13$\pm$130.40 & 0.00 & 659.68 & 5 \\
     
     \cellcolor{brown!20}10 & \cellcolor{brown!20}Text-to-SQL & 28   & 128 & 21.46$\pm$134.26 & 0.00 & 1284.67 & 391 & 40.20$\pm$87.74 & 0.00 & 850.28 & \cellcolor{gray!30}0 \\
     
     \cellcolor{brown!20}11 & \cellcolor{brown!20}Real Life Simulator & 28  & \cellcolor{blue!10}146 & 100.69$\pm$305.20 & 0.00 & 1803.46 & 365 & 44.88$\pm$90.68 & 0.01 & 707.10 & \cellcolor{gray!30}0 \\
     
     12 & Berachain GPT & 28   & 110 & 64.50$\pm$241.34 & 0.00 & 1739.92 & \cellcolor{blue!10}456 & 54.72$\pm$85.79 & 0.01 & 758.29 &\cellcolor{gray!30} 0 \\
     \hline
     
     13 & AI Research Agent & 32   & 67 & 13.98$\pm$46.38 & 0.00 & 302.88 & 413 & 30.03$\pm$57.75 & 0.01 & 527.84 & \cellcolor{blue!10}61345 \\

     \cellcolor{brown!20}14 & \cellcolor{brown!20}Move Code Co-creation & 14   & 11 & \cellcolor{gray!30}5.58$\pm$9.34 & 0.17 & 30.63 & 107 & \cellcolor{gray!30}13.06$\pm$39.90 & 0.00 & 269.48 & 14899 \\

     15 & Role-play LLM in Korean & 27  & 42 & 42.40$\pm$79.44 & 0.00 & 318.89 & 352 & 62.39$\pm$125.80 & 0.00 & 854.83 & 60830 \\

     16 & Investment Report Generator & 15  & 24 & 6.04$\pm$10.14 & 0.05 & 40.03 & 98 & 18.36$\pm$40.23 & 0.03 & 249.95 & 18770 \\
     \hline
     \multicolumn{2}{c|}{Total} & 224 & 603 & 165.11$\pm$403.45 & 0.00 & 3895.53 & 1051 & 69.80$\pm$158.77 & 0.00 & 2850.25 & 63265\\
        \bottomrule
    \end{tabular}
    }
    \caption{A total of 16 tasks have been trained and validated over a period of seven months. Starting from Task 13, participants are able to create delegation pools to create delegators, resulting in a significant increase in the number of delegators.}
    % \vspace{-8mm}
    \label{tab:task}
\end{table*}

\section{Implementation and Evaluation}

\subsection{On-Chain Implementation}
We implement \AIArena on the public Base blockchain Sepolia testnet. We leverage Solidity to implement the on-chain reward calculation and distribution smart contracts. We run the system from April 30, 2024, to December 9, 2024. For each day, $1074$ tokens will be minted and rewarded to the participants, and $\gamma = 0.7$. Table $1$ reports the summary statistics of all trained tasks on \AIArena.

% The reward ratio between training nodes and validators for tasks 1-2 was set to 7:3; between tasks 3-5, it was 6:4; it was 5:5 from tasks 6-8. From task 9 onwards, the ratio is no longer predefined; rather, it depends on the relative stake amount of training nodes and validators in a given task. 

% Note also that $\sigma$, which determines the reward ratio between training nodes or validators and their respective delegators, was set as 10\% for tasks 1-12. From task 13 onwards, training nodes and validators are free to set this reward ratio. 

% \vspace{-5mm}

\begin{figure}[H]
\centering
\includegraphics[width=\columnwidth]{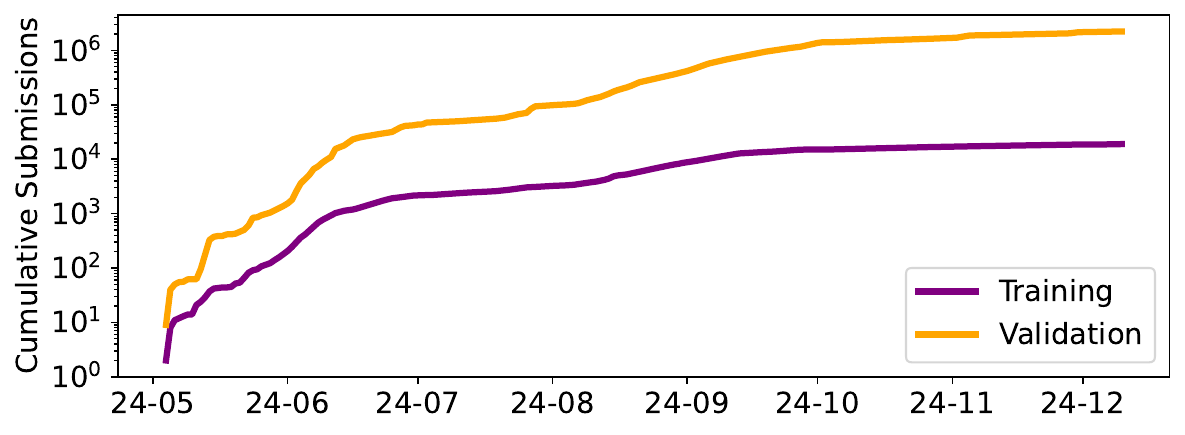}
\vspace{-8mm}
\caption{Training and validation submissions over time.}
\label{fig:ai_arena_submission_over_time}
\end{figure}

\noindent
\textbf{Number of Participants.} In total, we observe that $603$ training nodes, $1{,}051$ validators, and $63{,}265$ delegators have participated in the system to train and validate the $16$ various tasks, accumulating $18{,}926$ training and $2{,}225{,}254$ validation submissions (see Fig.~\ref{fig:ai_arena_submission_over_time}). Across most tasks, the number of validators consistently exceeds the number of training nodes.  For example, in task 10, overall $391$ validators participated, compared to only $128$ training nodes. Even for smaller-scale tasks such as task 14, the number of validators outnumbered training nodes.  

% \vspace{-5mm}

% \vspace{-5mm}
\noindent
\textbf{Reward Distribution.} The reward distribution among training nodes and validators reveals interesting patterns. Training nodes generally receive higher rewards per participant compared to validators before in tasks 1-12, due to the smaller number of training nodes. However, the variability in rewards among training nodes is much higher, as evidenced by the standard deviation, which indicates a broader range of effort and contribution levels. This suggests that while training tasks may offer higher potential rewards, they also come with greater uncertainty, whereas validation tasks provide more consistent but relatively lower rewards. Such dynamics could further explain the observed user preference for validation tasks, which offer steadier returns with less variance in earnings.

\begin{table}[t]
\centering
\renewcommand\arraystretch{1}
\resizebox{\linewidth}{!}{
\begin{tabular}{c|c|c|c|c|c}
\hline
\multirow{2}{*}{\textbf{LLM}} 
& \multicolumn{1}{|c|}{\makecell{{\colorbox{brown!20}{\textbf{Text-to-SQL }}} \\ (Task 10)}} 
& \multicolumn{1}{|c|}{\makecell{{\colorbox{brown!20}{\textbf{Life Simulator}}}\\(Task 11)}} 
& \multicolumn{3}{|c}{\makecell{\colorbox{brown!20}{\textbf{Code Co-creation }} \\(Task 14)}} \\ \cline{2-6}

~ & \textbf{$\mathcal{L}$} & \textbf{$\mathcal{L}$} & \textbf{$\mathcal{L}_\text{CMC}$} & \textbf{$\mathcal{L}_\text{MBLC}$} & \textbf{$\mathcal{L}_\text{N}$} \\ \hline
Rank-1 & \cellcolor{red!20}0.080 & \cellcolor{red!20}0.648 & - & - & \cellcolor{green!20}0.257 \\ 
% \hline
Rank-2 & 0.084 & \cellcolor{red!20}0.648 & - & - & 0.272 \\ 
% \hline
Rank-3 & 0.085 & 0.688 & - & - & 0.284 \\ 
% \hline
Phi3-4B* & 1.030 & 1.073 & 0.450 & 0.533 & 0.504 \\ 
% \hline
% Note: Phi3 is Phi-3-mini-4k-instruct
Qwen2.5-7B* & 1.055 & 1.397 & 0.379 & 0.530 & 0.575 \\ 
% \hline
% Note: Qwen2.5-7B* some of result is instruct version
Qwen2.5-14B** & 1.090 & 1.345 & 0.370 & 0.512 & 0.556 \\ \
% hline
Yi1.5-9B* & 0.322 & 1.288 & 0.631 & 0.726 & 0.631 \\ \hline
\end{tabular}
}
\caption{Model performance (loss $\mathcal{L}$) across three tasks. 
\\
\footnotesize{\textnormal{CMC and MBLC denote models finetuned exclusively on the respective datasets: CMC (Co-created Move Code from \AIArena) and MBLC (Move-Bytecode-LLVM-Compiler project from CMC). N indicates the native output of the model without any finetuning. \\
*: Models based on foundational LLMs;
**: SOTA, larger-scale LLMs.}}}
% \vspace{-10mm}
\label{tab:all_tasks}
\end{table}

\subsection{Off-Chain Implementation}
% Contributions
% 1. Co-created code dataset （Task 14 Aptos）
% 2. Finetuned model (1. Task 9 Text2SQL 2, Task 11, Life Simulator )

\begin{figure}[t]
    \centering
    \includegraphics[width=\linewidth]{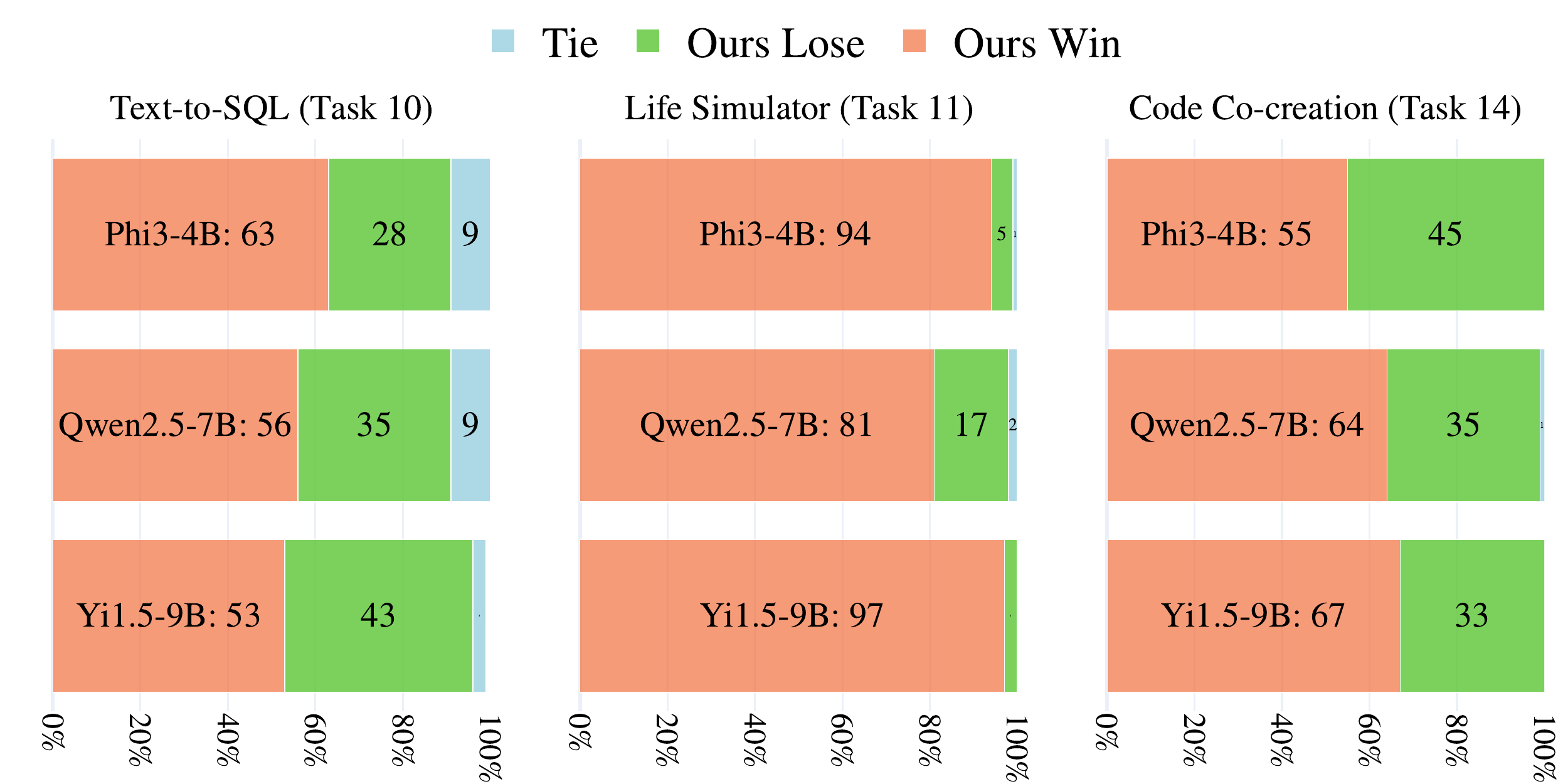}
    \caption{Comparison of the Rank-1 Model in \AIArena with Baselines, evaluated using the Human Preferences Evaluation method judged by GPT-4o.}
    \label{fig:compair}
\end{figure}

To demonstrate the feasibility of \AIArena in realistic scenarios, we evaluated three diverse and popular tasks using the methodology proposed by the platform\footnote{\url{https://github.com/FLock-io/llm-loss-validator}}. The results are summarized in Table~\ref{tab:all_tasks}. Meanwhile, we employ the Human Preferences Evaluation method, with the voting results by GPT-4o presented in Fig.~\ref{fig:compair}. The results show that all Rank-1 training nodes’ models outperform baseline models, larger SOTA models, and baseline LLMs finetuned on the provided or subset CMC and MBLC datasets.

% Text-to-SQL is a specialized NLP task where the objective is to translate natural language descriptions into SQL queries. While this approach has made strides in various domains, it remains largely unexplored for complex, real-world blockchain data. Such capabilities are crucial for enabling domain experts, analysts, and even non-technical stakeholders to automatically query and interpret on-chain transaction records, token transfers, and smart contract states, ultimately fostering more accessible and data-driven blockchain insights.

% To advance this area, we introduced a text-to-SQL challenge on the AI Arena platform, starting from a seed dataset constructed by structuring previously unorganized blockchain transaction data. Participants were encouraged to diversify and enrich this dataset by adding new examples and refining schema representations, thereby creating a more comprehensive resource. We then fine-tuned LLMs on these enriched samples to enhance their text-to-SQL skills specifically for blockchain queries. 
\noindent
\colorbox{brown!20}{\textbf{Text-to-SQL for Onchain Data (Task 10).}}
Text-to-SQL, which translates natural language into SQL queries, has advanced in various domains but remains largely unexplored for complex blockchain data, a gap that limits accessible, data-driven insights into on-chain transactions, token transfers, and smart contract states. To address this problem, we launched a text-to-SQL task on the \AIArena platform, starting from a seed dataset of structured blockchain transaction data, and encouraged participants to enrich it with diverse examples and refined schemas. All participants then fine-tuned LLMs on these enhanced samples, thereby improving their text-to-SQL capabilities specifically for blockchain queries.

\noindent
\colorbox{brown!20}{\textbf{Real Life Simulator (Task 11).}}
Life Simulator is a dynamic, LLM-powered game where players guide a character’s growth through career, relationships, and personal choices, experiencing unique, adaptive outcomes in a realistic virtual world. However, most current LLMs tend to generate overly optimistic life stories, which deviate significantly from realism and undermine the core purpose of a life simulator. To address this, the \AIArena platform has been utilized to gather contributions from the community, enhancing LLMs’ ability to generate more realistic and diverse life scenarios.

\noindent
\colorbox{brown!20}{\textbf{Code Co-creation (Task 14).}}
Accurate code generation and understanding are critical when leveraging LLMs for advanced automation, particularly in low-resource blockchain languages. Building on this need, we focus our efforts on the Move language, aiming to develop a model capable of generating accurate instructions and comments. Through our \AIArena platform, a diverse community of contributors collectively curated what we believe, to the best of our knowledge, to be the largest Move instruction dataset to date (see {https://huggingface.co/datasets/flock-io/move-code-comment}). Drawing from 514 distinct Move projects, they enriched the source code with detailed instructions and comments.

\section*{Acknowledgements}
This work was partially supported by a research grant from the Ethereum Foundation.

\bibliographystyle{ACM-Reference-Format}
\balance
\bibliography{references}


\begin{thebibliography}{11}

%%% the Web address.

\ifx \showCODEN    \undefined \def \showCODEN     #1{\unskip}     \fi
\ifx \showDOI      \undefined \def \showDOI       #1{#1}\fi
\ifx \showISBNx    \undefined \def \showISBNx     #1{\unskip}     \fi
\ifx \showISBNxiii \undefined \def \showISBNxiii  #1{\unskip}     \fi
\ifx \showISSN     \undefined \def \showISSN      #1{\unskip}     \fi
\ifx \showLCCN     \undefined \def \showLCCN      #1{\unskip}     \fi
\ifx \shownote     \undefined \def \shownote      #1{#1}          \fi
\ifx \showarticletitle \undefined \def \showarticletitle #1{#1}   \fi
\ifx \showURL      \undefined \def \showURL       {\relax}        \fi
% invisible to TeX
\providecommand\bibfield[2]{#2}
\providecommand\bibinfo[2]{#2}
\providecommand\natexlab[1]{#1}
\providecommand\showeprint[2][]{arXiv:#2}

\bibitem[Challen et~al\mbox{.}(2019)]%
        {challen2019artificial}
\bibfield{author}{\bibinfo{person}{Robert Challen}, \bibinfo{person}{Joshua Denny}, \bibinfo{person}{Martin Pitt}, \bibinfo{person}{Luke Gompels}, \bibinfo{person}{Tom Edwards}, {and} \bibinfo{person}{Krasimira Tsaneva-Atanasova}.} \bibinfo{year}{2019}\natexlab{}.
\newblock \showarticletitle{Artificial intelligence, bias and clinical safety}.
\newblock \bibinfo{journal}{\emph{BMJ quality \& safety}} \bibinfo{volume}{28}, \bibinfo{number}{3} (\bibinfo{year}{2019}), \bibinfo{pages}{231--237}.
\newblock


\bibitem[Chen et~al\mbox{.}({[n.\,d.]})]%
        {chen2018machine}
\bibfield{author}{\bibinfo{person}{Xuhui Chen}, \bibinfo{person}{Jinlong Ji}, \bibinfo{person}{Changqing Luo}, \bibinfo{person}{Weixian Liao}, {and} \bibinfo{person}{Pan Li}.} \bibinfo{year}{[n.\,d.]}\natexlab{}.
\newblock \showarticletitle{When machine learning meets blockchain: A decentralized, privacy-preserving and secure design}. In \bibinfo{booktitle}{\emph{2018 IEEE Big Data}}.
\newblock


\bibitem[Dong et~al\mbox{.}(2024)]%
        {dong2024defending}
\bibfield{author}{\bibinfo{person}{Nanqing Dong}, \bibinfo{person}{Zhipeng Wang}, \bibinfo{person}{Jiahao Sun}, \bibinfo{person}{Michael Kampffmeyer}, \bibinfo{person}{William Knottenbelt}, {and} \bibinfo{person}{Eric Xing}.} \bibinfo{year}{2024}\natexlab{}.
\newblock \showarticletitle{Defending against poisoning attacks in federated learning with blockchain}.
\newblock \bibinfo{journal}{\emph{IEEE Transactions on Artificial Intelligence}} (\bibinfo{year}{2024}).
\newblock


\bibitem[Etzioni(2016)]%
        {Etzioni2016Democratizing}
\bibfield{author}{\bibinfo{person}{Oren Etzioni}.} \bibinfo{year}{2016}\natexlab{}.
\newblock \showarticletitle{Democratizing artificial intelligence through open AI models}.
\newblock \bibinfo{journal}{\emph{AI Magazine}} \bibinfo{volume}{37}, \bibinfo{number}{1} (\bibinfo{year}{2016}), \bibinfo{pages}{50--52}.
\newblock


\bibitem[Jia et~al\mbox{.}({[n.\,d.]})]%
        {jia2021proof}
\bibfield{author}{\bibinfo{person}{Hengrui Jia}, \bibinfo{person}{Mohammad Yaghini}, \bibinfo{person}{Christopher~A Choquette-Choo}, \bibinfo{person}{Natalie Dullerud}, \bibinfo{person}{Anvith Thudi}, \bibinfo{person}{Varun Chandrasekaran}, {and} \bibinfo{person}{Nicolas Papernot}.} \bibinfo{year}{[n.\,d.]}\natexlab{}.
\newblock \showarticletitle{Proof-of-learning: Definitions and practice}. In \bibinfo{booktitle}{\emph{2021 IEEE S\&P}}.
\newblock


\bibitem[Lan et~al\mbox{.}(2021)]%
        {lan2021proof}
\bibfield{author}{\bibinfo{person}{Yixiao Lan}, \bibinfo{person}{Yuan Liu}, \bibinfo{person}{Boyang Li}, {and} \bibinfo{person}{Chunyan Miao}.} \bibinfo{year}{2021}\natexlab{}.
\newblock \showarticletitle{Proof of learning (PoLe): Empowering machine learning with consensus building on blockchains}. In \bibinfo{booktitle}{\emph{AAAI}}.
\newblock


\bibitem[Montes and Goertzel(2019)]%
        {montes2019distributed}
\bibfield{author}{\bibinfo{person}{Gabriel~Axel Montes} {and} \bibinfo{person}{Ben Goertzel}.} \bibinfo{year}{2019}\natexlab{}.
\newblock \showarticletitle{Distributed, decentralized, and democratized artificial intelligence}.
\newblock \bibinfo{journal}{\emph{Technological Forecasting and Social Change}} (\bibinfo{year}{2019}).
\newblock


\bibitem[Salah et~al\mbox{.}(2019)]%
        {salah2019blockchain}
\bibfield{author}{\bibinfo{person}{Khaled Salah}, \bibinfo{person}{M~Habib~Ur Rehman}, \bibinfo{person}{Nishara Nizamuddin}, {and} \bibinfo{person}{Ala Al-Fuqaha}.} \bibinfo{year}{2019}\natexlab{}.
\newblock \showarticletitle{Blockchain for AI: Review and open research challenges}.
\newblock \bibinfo{journal}{\emph{IEEE access}} (\bibinfo{year}{2019}).
\newblock


\bibitem[Wang et~al\mbox{.}(2024)]%
        {wang2024sok}
\bibfield{author}{\bibinfo{person}{Zhipeng Wang}, \bibinfo{person}{Rui Sun}, \bibinfo{person}{Elizabeth Lui}, \bibinfo{person}{Vatsal Shah}, \bibinfo{person}{Xihan Xiong}, \bibinfo{person}{Jiahao Sun}, \bibinfo{person}{Davide Crapis}, {and} \bibinfo{person}{William Knottenbelt}.} \bibinfo{year}{2024}\natexlab{}.
\newblock \showarticletitle{SoK: Decentralized AI (DeAI)}.
\newblock \bibinfo{journal}{\emph{arXiv preprint arXiv:2411.17461}} (\bibinfo{year}{2024}).
\newblock


\bibitem[Zhang et~al\mbox{.}(2020)]%
        {zhang2020model}
\bibfield{author}{\bibinfo{person}{Jie Zhang}, \bibinfo{person}{Dongdong Chen}, \bibinfo{person}{Jing Liao}, \bibinfo{person}{Han Fang}, \bibinfo{person}{Weiming Zhang}, \bibinfo{person}{Wenbo Zhou}, \bibinfo{person}{Hao Cui}, {and} \bibinfo{person}{Nenghai Yu}.} \bibinfo{year}{2020}\natexlab{}.
\newblock \showarticletitle{Model watermarking for image processing networks}. In \bibinfo{booktitle}{\emph{AAAI}}.
\newblock


\bibitem[Zou et~al\mbox{.}(2019)]%
        {zou2019smart}
\bibfield{author}{\bibinfo{person}{Weiqin Zou}, \bibinfo{person}{David Lo}, \bibinfo{person}{Pavneet~Singh Kochhar}, \bibinfo{person}{Xuan-Bach~Dinh Le}, \bibinfo{person}{Xin Xia}, \bibinfo{person}{Yang Feng}, \bibinfo{person}{Zhenyu Chen}, {and} \bibinfo{person}{Baowen Xu}.} \bibinfo{year}{2019}\natexlab{}.
\newblock \showarticletitle{Smart contract development: Challenges and opportunities}.
\newblock \bibinfo{journal}{\emph{IEEE TSE}} (\bibinfo{year}{2019}).
\newblock


\end{thebibliography}

\end{document}